\documentclass[12pt]{article}
\usepackage{epsfig}
\begin{document}
\begin{center}
{\Large {\bf $Z_6$ symmetry, electroweak transition, and magnetic monopoles at
high temperature}}

\vskip-40mm \rightline{\small ITEP-LAT/2006-04} \vskip 30mm

{%\baselineskip=16pt
\vspace{1cm}
{B.L.G.~Bakker$^a$, A.I.~Veselov$^b$, M.A.~Zubkov$^b$ }\\
\vspace{.5cm}
{\it $^a$ Department of Physics and Astronomy, Vrije Universiteit, Amsterdam,
The Netherlands \\
$^b$ ITEP, B.Cheremushkinskaya 25, Moscow, 117259, Russia }}
\end{center}

\begin{abstract}
We consider the lattice realization of the Standard Model with an additional
$Z_6$ symmetry. Numerical simulations were performed on the asymmetric lattice,
which corresponds to the finite temperature theory.  Our choice of parameters
corresponds to large Higgs masses  ($M_H > 90$ Gev). The phase diagram was
investigated and has been found to be different from that of the usual lattice
realization of the Standard Model. It has been found, that the
confinement-deconfinement phase transition lines for the $SU(2)$ and $SU(3)$
fields coincide. The transition line between Higgs and symmetric deconfinement
parts of the phase diagram and the confinement-deconfinement transition line
meet in a triple point. The transition between Higgs and symmetric parts of the
phase diagram corresponds to the finite temperature electroweak
transition/crossover. We see for the first time evidence that Nambu monopoles
are condensed at $T>T_c$ while at $T<T_c$ their condensate vanishes.
\end{abstract}

%\pacs{12.15.-y, 11.15.Ha, 12.10.Dm}

Owing to the present triviality bound the perturbation expansion in the
electroweak sector of the Standard Model is considered to work
perfectly at the energies of the electroweak scale $M_Z$ for Higgs
masses up to  $1$ Tev \cite{M_H}.  However, it was shown \cite{M_W_T},
that the finite temperature perturbation expansion breaks down at the
temperatures above the electroweak transition/crossover already for
Higgs masses above about $60$ GeV. Therefore the present lower bound on
the Higgs mass, $M_H > 114$ GeV, requires the use of nonperturbative
techniques while investigating electroweak physics at high
temperature.  Thus our understanding of electroweak physics, which is
based mainly on perturbation theory could be changed drastically at
temperatures close to and above the electroweak transition/crossover.

It was shown recently \cite{BVZ2003,BVZ2004, BVZ2005} that there is a
hidden $Z_6$ symmetry in the Higgs and fermion sectors of the Standard
Model.  The Standard Model on the lattice can be defined in such a way,
that the whole model is $Z_6$ invariant. The resulting model has the
same perturbation expansion as the usual lattice realization of the
Standard Model, which does not respect the $Z_6$ symmetry. On the other
hand, it was argued, that nonperturbatively those two lattice models
may represent different physics due to their different symmetry
properties.  In particular, it was supposed, that these two models may
describe the physics at temperatures close to the electroweak crossover in
different ways.

In this paper we report results of our investigation of the $Z_6$ symmetric
lattice version of the Standard Model at finite temperature. We considered the
model in the London limit, i.e. with infinite bare Higgs mass. This does not
mean, however, that the renormalized mass of the Higgs boson is
infinite\cite{Montvay}.

The phase diagram of our lattice model (Fig.~$1$) differs drastically from that
of the usual lattice realization of the Standard Model. Namely, in the latter
only one phase is present, the phase transition lines degenerate and become
crossover lines\cite{EW_T}.  Our lattice model clearly contains three parts.
The first one (I) is the confinement phase, where the $SU(3)$ fields confine
quarks, and confinement-like forces are observed between the leptons. In the
next one (II) there are no confinement-like forces at all but the line-like
objects which arise in the unitary gauge are found to be condensed. We identify
these objects with quantum generalization of the well-known classical Nambu
monopole configurations\cite{Nambu}. The last part of the phase diagram (III)
corresponds to the low temperature physics, where the Higgs field is condensed.
In this part of the phase diagram Nambu monopoles are not condensed and their
density is dropping rapidly when moving away from the transition line. Further
we refer to both parts II and III of the diagram as to phases, taking in mind,
however, that the transition line between them could be actually a crossover
line.

The increase of temperature corresponds to a shift from phase III to phase II.
The physical temperature is expressed here as $T = \frac{1}{a N_T}$, where
$N_T$ is the time extent of the lattice while $a$ is the lattice spacing, which
depends on the values of the coupling constants. We denote the value of
temperature at the transition point as $T_c$. Here the monopole condensate
plays the role of an order parameter. We do not observe a vanishing of the
Nambu monopole condensate in phase III with increasing lattice size. This
result leads us to suggest the hypothesis, that those monopoles survive and are
condensed in the continuum theory at $T>T_c$. This is in accordance with the
supposition which was made in the framework of the $SU(2)$ Higgs model in
\cite{Chernodub_Nambu} \footnote{In \cite{Chernodub_Nambu} it was shown that at
a certain limit of the coupling constants the $SU(2)$ Higgs model becomes
identical to the Georgi-Glashow model. Then, 't Hooft-Polyakov monopoles were
identified with Nambu monopoles. Therefore, condensation of 't Hooft-Polyakov
monopoles in the symmetric phase of the Georgi-Glashow model means that at
least in the limit of coupling constants considered the Nambu monopoles are
condensed in the symmetric phase of the $SU(2)$ Higgs model.}. It was argued in
\cite{EW_T} that in the usual definition of the lattice Standard Model the
electroweak transition is actually a crossover at the allowed values of the
Higgs mass. This has led to the conclusion that the baryon asymmetry could not
be produced during the electroweak phase transition, as was suggested in
\cite{Rubakov}. Our investigation shows, however, that the vacuum structure
below and above the transition is different. Therefore, we do not exclude that
there is a phase transition of a high order at $T=T_c$, where the condensate of
electromagnetic monopoles vanishes. Although the high order phase transitions
are not well understood, they are known to exist in some lattice and
statistical systems \cite{Janke_T}. The situation may also be similar to that
of the $3D$ Compact $U(1)$ Lattice Higgs Model \cite{Janke}, where the phase
boundary consists of a line of first-order phase transitions at small Higgs
self-coupling, ending at a critical point. The phase boundary then continues as
a Kertesz line across which thermodynamic quantities are nonsingular. It is
worth mentioning, that within the $3D$ $SU(2)$ Higgs model \cite{Chernodub} it
was found that the Z-vortices percolate at $T>T_c$ while at $T<T_c$ they do
not. For this reason in Ref.~\cite{Chernodub} this transition was called the
``percolation transition". It is also in accordance with our observations, as
in $4D$ Z-vortices are known to terminate at Nambu monopoles.

We consider here the lattice model described in \cite{BVZ2005}.  We use
asymmetric lattices with time extents $2$ and $4$ and of space sizes from $8^3$
up to  $24^3$. With the definitions of \cite{BVZ2005} the pure gauge part of
the action has the form
\begin{eqnarray}
 S_g & = & \beta \sum_{\rm plaquettes}  \{2(1-\mbox{${\small
 \frac{1}{2}}$} {\rm Tr}\, U_p \cos \theta_p) +
\nonumber \\
 && +\;(1-\cos 2\theta_p)
\nonumber \\
 && +\;6[1-\mbox{${\small \frac{1}{6}}$} {\rm Re Tr}  \,
 \Gamma_p {\rm Tr}\, U_p {\rm exp} (i\theta_p/3)]
\nonumber\\
 && +\;3[1-\mbox{${\small \frac{1}{3}}$} {\rm Re Tr}  \,
 \Gamma_p {\rm exp} (-2i\theta_p/3)]
\nonumber \\
 && +\;3[1-\mbox{${\small \frac{1}{3}}$} {\rm Re Tr}  \,
 \Gamma_p {\rm exp} (4i\theta_p/3)]\},
\label{Act}
\end{eqnarray}
where the sum runs over the elementary plaquettes of the lattice, and
\begin{eqnarray}
 \Gamma \in SU(3), \quad U \in SU(2), \quad e^{i\theta} \in U(1).
\end{eqnarray}
Each term of the action, Eq.~(\ref{Act}), corresponds to a parallel
transporter along the boundary  $\partial p$ of a plaquette $p$.

The action for the scalar field is considered in its simplest form
\cite{BVZ2004} in the London limit, i.e., in the limit of infinite bare Higgs
mass. After fixing the unitary gauge we obtain:
\begin{eqnarray}
 S_H & = & \gamma \sum_{xy}[1 - Re(U^{11}_{xy} e^{i\theta_{xy}})].
\end{eqnarray}
Here $\gamma = v^2$, where $v$ is the bare scalar field vacuum
average.  We consider our model in quenched approximation, i.e., we
neglect the effect of virtual fermion loops. Thus the whole action of
the model is $S = S_g + S_H$.

It is worth mentioning, that the action of the form (\ref{Act})
actually appears as a low energy approximation of the $SU(5)$
GUT\cite{BVZ2003}.  Therefore, the bare coupling $\beta$ (which is the
same for all terms of the action) could be considered at the GUT
scale.  Then, owing to the renormalization group equations, the
renormalized gauge couplings at the electroweak scale $M_Z$ come close
to the experimental ones and coincide with them up to a few percents.
Actually, their change is of logarithmic form and is very slow.  The
physical scale (i.e. the value of the lattice spacing in physical
units) should be determined using a measurement of the zero-temperature Z-boson
mass in lattice units. We did not perform extensive calculation of $M_Z$ in our
model within the considered ranges of bare couplings. However, our measurements
of correlators in the vector channel allow us to evaluate the lattice spacing
in the considered region of phase III ($\beta \in (0.6,0.8)$, $\gamma \in (1.0,
1.5)$) to be  $(130\pm 40\, {\rm GeV})^{-1}$.

The renormalized $\alpha_s$ (which stands for the strong interaction of
quarks) in our model can be calculated using certain correlators of
colored fields and should be expressed as a function of bare couplings.
In principle, if we start from the theory at the GUT scale, such a
calculation must give a reasonable result.  However, due to the
technical problems in lattice simulations, we expect direct
calculations at the Electroweak scale would give unphysically small
values of $\alpha_s(M_Z)$. This means that at the energies of the order
of $100$ GeV color fields in our lattice model appear to be suppressed.
Thus we consider their influence on the Electroweak dynamics only
qualitatively. However, if such an influence (which is due to the
specific $Z_6$ invariant terms in the action) is found, it is
reasonable to expect that it also should take place for the realistic
case of unsuppressed colored fields.

In general all the renormalized gauge couplings should be calculated
using static quark and lepton potentials.  Then the lines of constant
physics (LCP) in the space of bare parameters (including time extent of
the lattice) are defined as the lines, where the zero-temperature
renormalized couplings and the temperature in physical units are
constant\cite{Montvay,EW_T}. In the present paper we do not consider
LCP, which is needed for a proper investigation of approaching continuum
physics\footnote{We must notice here that for the proper investigation
of approaching the continuum physics along the LCP it could become
necessary to extend the space of bare couplings and include there
different gauge couplings for $SU(2)$, $SU(3)$, $U(1)$ fields as well
as the bare Higgs boson mass.}.
The consideration of LCP is also necessary for determining the
correspondence between the phase diagram in the $\beta$ - $\gamma$
plane and the conventional phase diagram in the $T - M_H$ plane.
However, we can understand the emergence of temperature in the phase
diagram represented in the Fig. 1 using naive expressions for the lattice
spacing and the Z-boson mass. Namely, in phase III at tree level
$M_Z = g_z v/2$  in lattice units. Here, $g_z = g/{\rm cos} \,
\theta_W$, $8 \beta = \frac{4}{g^2}$, and ${\rm cos}^2 \, \theta_W =
\frac{5}{8}$.  Therefore $M_Z \sim \sqrt{\frac{\gamma}{5\beta}}$.
Next, the lattice spacing in physical units is equal to $a = M_Z /M^{\rm
phys}_Z$, where $M^{\rm phys}_Z$ is about $90$ GeV. Therefore, say, at
$\beta = 0.6$, $\gamma = 1.5$ the naive tree level estimate for the lattice
spacing is $a \sim (130\, {\rm GeV})^{-1}$, while at $\beta = 0.8$,
$\gamma = 1.0$ it is  $a \sim (180\, {\rm GeV})^{-1}$. Finally, the
temperature is estimated as $T = \frac{1}{N_Ta} \sim
\sqrt{\frac{5\beta}{\gamma N_T^2}}\, 90\, {\rm GeV}$. Although the last
expression is to be modified using the lattice renormalization group
equations, it shows, that, in general, temperature is increased with
an increase of $\beta$ and a decrease of $\gamma$ and $N_T$.

Physically interesting values of the coupling constants could be
evaluated following the naive estimates considered above. In our model
the bare electromagnetic charge is $e^2 = g^2 {\rm sin}^2 \, \theta_W =
\frac{3}{16 \beta}$; the experimental value is $\frac{e^2(M_Z)}{4\pi}\sim
\frac{1}{128}$. Thus $\beta \sim \frac{6}{\pi} \sim 1.9$. Our estimate
for the critical $\gamma$ at $\beta = 1.9$ is $\gamma_c \sim 0.9$.
Therefore, the critical temperature could be estimated as $T_c \sim 150$
 GeV. Of course, this is a very rough estimate and it should be
improved using direct lattice methods\footnote{This value should be compared
with the one calculated within the $SU(2)$ gauge Higgs model in \cite{Rum}.
There for $M_H = 120$ GeV $T_c$ was found to be of the order of $210$ GeV while
for $M_H = 180$ GeV $T_c \sim 250$ GeV}. For technical reasons we did not
perform extensive numerical simulations in the vicinity of $\beta \sim 1.9$.
Instead we investigated the region $\beta \in (0.6, 0.8)$ of phase III. This
means that our results at the present moment should be considered as
qualitative only.

The following variables are considered as creating a $Z$ boson and a $W$ boson,
respectively:
\begin{eqnarray}
  Z_{xy} & = & Z^{\mu}_{x} \;
 = {\rm sin} \,[{\rm Arg} U_{xy}^{11} + \theta_{xy}],
\nonumber\\
 W_{xy} & = & W^{\mu}_{x} \,= \,U_{xy}^{12} e^{i\theta_{xy}}.
\end{eqnarray}
Here, $\mu$ represents the direction $(xy)$.

After fixing the unitary gauge the electromagnetic $U(1)$ symmetry remains:
\begin{eqnarray}
 U_{xy} & \rightarrow & g^\dag_x U_{xy} g_y, \nonumber\\
 \theta_{xy} & \rightarrow & \theta_{xy} -  \alpha_y/2 + \alpha_x/2,
\end{eqnarray}
where $g_x = {\rm diag} (e^{i\alpha_x/2},e^{-i\alpha_x/2})$.

In the unitary gauge there is also a $U(1)$ lattice gauge field, which is
defined as
\begin{equation}
 A_{xy}  =  A^{\mu}_{x} \;
 = \,[-{\rm Arg} U_{xy}^{11} + \theta_{xy}]  \,{\rm mod} \,2\pi,
\end{equation}

The fields $A$, $Z$, and $W$ transform as follows:
\begin{eqnarray}
 A_{xy} & \rightarrow & A_{xy} - \alpha_y + \alpha_x, \nonumber\\
 Z_{xy} & \rightarrow & Z_{xy}, \nonumber\\
 W_{xy} & \rightarrow & W_{xy}e^{-i\alpha_x}.
\label{T}
\end{eqnarray}

It should be mentioned that the field $A$ cannot be treated as a usual
electromagnetic field as the set of variables $A$, $Z$, and $W$ do not
diagonalize the kinetic part of the pure gauge action (\ref{Act}) in
its naive continuum limit.  In our lattice model the electromagnetic
field $A_{\rm em}$ should be defined as
\begin{equation}
 A_{\rm em}  =  A + Z^{\prime} - 2 \,{\rm sin}^2\, \theta_W Z^{\prime},
\label{A_em}
\end{equation}
where $Z^{\prime} = [{\rm Arg} U_{xy}^{11} + \theta_{xy}]{\rm mod}
2\pi$. The naive value of the Weinberg angle corresponds to ${\rm
sin}^2 \, \theta_W = \frac{3}{8}$. However, the renormalized Weinberg
angle is to be calculated through the ratio of the lattice masses:
${\rm cos} \, \theta_W = \frac{M_W}{M_Z}$.

In order to evaluate the zero temperature masses of the $Z$-boson and Higgs
boson we use the correlators:
\begin{eqnarray}
 \langle \sum_{\mu} Z^{\mu}_{x} Z^{\mu}_{y} \rangle  & \sim &
 \frac{1}{|x-y|^2} e^{-M_{Z}|x-y|}\nonumber\\
 \langle\langle H_{x} H_{y}\rangle\rangle  & = &
 \langle H_{x} H_{y} \rangle - \langle H \rangle^2  \sim
 \frac{1}{|x-y|^2}  e^{-M_{H}|x-y|},\label{cor}
\end{eqnarray}
where  $H_x = \sum_{y} |W_{xy}|^2$ or $H_x =  \sum_{y} Z^2_{xy}$.

The position of the transition lines on the phase diagram of the finite
temperature model almost coincide with that of the lines on the phase
diagram of the zero temperature model\cite{BVZ2005}. The only
difference is that the transition line between phases II and III is
shifted to higher values of $\gamma$. Our statistics does not allow us
to perform a precise calculation of the values of $M_H$ and $M_Z$.
Therefore we have made a very rough estimate of their ratio. Namely, in
phase III of the zero temperature model (on the lattice $16^4$) for the
considered values of the couplings ($\beta \in (0.5, 0.8)$, and
$\gamma$ up to $1.5$) our estimate is $M_H/M_Z \sim 2.5\pm 0.5$. This
is in qualitative agreement with the predictions made within the
$SU(2)$ Higgs model considered in the London limit \cite{Montvay}. Thus
in our model the estimate for the Higgs mass could  be $M_H \sim 230
\pm 50$ GeV. However, as it was mentioned above, the lattice spacing in
the considered part of the phase diagram is estimated to be  $(130 \pm
40\,{\rm GeV})^{-1}$. In lattice theory the inverse lattice spacing
plays the role of an ultraviolet cutoff. In general, quantum field
theory does not work at energies higher than the ultraviolet cutoff.
Therefore, we feel it necessary to weaken our estimate for $M_H$.
Namely, we evaluate it as $M_H>90$ GeV.

To understand the dynamics of external charged particles, we consider
the Polyakov lines defined on the asymmetric lattice in the fermion
representations listed in the table in \cite{BVZ2005}:
\begin{eqnarray}
 {\cal P}^{\rm L}_{\rm lept} & = &
 \langle {\rm Re} {\rm Tr} \,\Pi_{(xy) \in l} U_{xy}  e^{-i\theta_{xy}}\rangle, \nonumber\\
 {\cal P}^{\rm R}_{\rm lept} & = & \langle {\rm Re} \Pi_{(xy) \in l} \,
 e^{-2i\theta_{xy}}\rangle ,
\nonumber\\
 {\cal P}^{\rm L}_{ {\rm quarks}} & = & \langle {\rm Re} \Pi_{(xy) \in l}  \,
 \Gamma_{xy} \, U_{xy}\, e^{\frac{i}{3}\theta_{xy}}\rangle ,
\nonumber\\
 {\cal P}^{\rm R}_{{\rm down} \, {\rm quarks}} & =  &
 \langle {\rm Re} \Pi_{(xy) \in l} \, \Gamma_{xy} \,
 e^{-\frac{2i}{3}\theta_{xy}}\rangle ,
\nonumber\\
 {\cal P}^{\rm R}_{{\rm up} \, {\rm quarks}} & = &
 \langle {\rm Re} \Pi_{(xy) \in l} \, \Gamma_{xy} \,
 e^{\frac{4i}{3}\theta_{xy}}\rangle .
\label{WL}
\end{eqnarray}
Here $l$ denotes a line on the lattice in the time direction, which is
closed due to the periodic boundary conditions.

It is found that to the left of the vertical line of the phase diagram
all Poliakov lines vanish while to the right of this line all of them
increase rapidly (Fig.~$2$).

In order to extract physical information from the $SU(3)$ fields themselves in
a particulary simple way we use the so-called indirect Maximal Center
Projection (see, for example, \cite{Greensite,BVZ1999}).

We investigated several types of monopoles.  The monopoles, which carry
information about colored fields are extracted from the composite fields $C^i$
(for their definition see \cite{BVZ2005}) constructed of the $SU(2)$ and $U(1)$
fields and of the center vortices appearing in the Maximal Center Projection of
the color group:
\begin{equation}
 j_{C^i} =  \frac{1}{2\pi} {}^*d([d C^i]{\rm mod}2\pi) .
\end{equation}
(Here we used the notations of differential forms on the lattice. For a
definition of those notations see, for example, ~\cite{forms}.) Pure
$U(1)$ monopoles, corresponding to the second term in (\ref{Act}), are
extracted in the same way from $2\theta$: $j_{2\theta} = \frac{1}{2\pi}
{}^*d([d 2\theta]{\rm mod}2\pi)$. We refer to them as hypercharge
monopoles.

The electromagnetic monopoles must be related to the field $A_{\rm
em}$. However, $A_{\rm em}$ itself is not a usual lattice $U(1)$ field
that should be periodic with the period $2\pi$. Instead $A_{\rm em}$ is
constructed of the two $U(1)$ variables: $A$ and $Z^{\prime}$.
Therefore, the electromagnetic monopoles should be constructed of
either $A$ or $Z^{\prime}$ fields, or, possibly both of them.
Therefore, we denote these monopoles by $j_{A} = \frac{1}{2\pi}
{}^*d([d A]{\rm mod}2\pi)$ and $j_{Z} = \frac{1}{2\pi} {}^*d([d
Z^{\prime}]{\rm mod}2\pi)$.

The worldsheet of the quantum $Z$-string may be defined as $\sigma_Z =
\frac{1}{2\pi}\{ [d Z^{\prime}]\, {\rm mod}2\pi - d Z^{\prime} \} $.
This is actually a Nielsen-Olesen string embedded in the Standard
Model.

On the classical level the singularity of the hypercharge field
$2\theta = [A+Z^{\prime}]{\rm mod} 2\pi$ is suppressed by the $U(1)$
pure gauge field action. Therefore, one would expect that $j_Z = -
j_A$. This situation corresponds to the appearance of the quantum Nambu
monopole with the worldline $j_A$. From (\ref{A_em}) it follows that
its magnetic charge is proportional to $4\pi{\rm sin}^2\theta_W$ as it
should \cite{Nambu}.

In lattice quantum theory, however, the singularities of the
hypercharge field may appear in the form of the corresponding
monopoles. This situation corresponds to the appearance of
$j_{2\theta}$. Then in the absence of a $Z$-string the magnetic charge
of such  configurations appear to be proportional to $2\pi$. This case
of the magnetic monopole for even $j_A$ seems to be corresponding to a
Cho-Maison monopole or dyon \cite{Cho} \footnote{Another way to
understand the appearance of a Cho-Maison monopole with the magnetic
charge $4\pi$ is to consider the monopole current extracted from the
field $\theta$. Then, due to the identity $A_{\rm em} = 2 \theta -
2{\rm sin}^2 \theta_W Z^{\prime}$ the magnetic  charge of such a
monopole current is $4\pi$ in the absence of the $Z$-string.}.

Thus, we arrive at the following two possibilities:

1. In the absence of $j_{2\theta}$ the A-monopoles $j_A$ represent
Nambu monopoles with the magnetic charge $4\pi{\rm sin}^2\theta_W$.

2. If $j_{2\theta} \ne 0$ the A-monopoles may represent another type of
monopoles with the magnetic charge $2\pi$. Such a monopole with even
$j_A$ corresponds to Cho-Maison monopole or dyon.

The density of the monopoles is defined as follows:
\begin{equation}
 \rho = \left\langle \frac{\sum_{\rm links}|j_{\rm link}|}{4L^4}  \right\rangle,
\end{equation}
where $L$ is the lattice size.

It is found that the densities of the color and hypercharge monopoles
decrease rapidly to the right of the  vertical line in the phase
diagram and drop to zero soon after the phase transition. This,
together with the behavior of the Polyakov lines allows us to identify
the vertical line of the phase diagram with the
confinement-deconfinement phase transition common for $U(1)$, $SU(2)$,
and $SU(3)$ fields. The behavior of such quantities as the overall
action and the monopole densities possess hysteresis effects, which
allow us to suppose that this phase transition is of the first order.
The position of the phase transition is localized using hysteresis of
the action and is defined as the point where the mentioned Polyakov
lines vanish.

The monopole density $\rho(A)$ constructed of $j_A$ is found to be
nonzero for all values of the couplings considered within phase II. As
the hypercharge monopoles disappear in this phase, we identify here
$j_A$ with Nambu monopoles.  When going to phase III the density
decreases and vanishes soon after the transition. In order to
investigate the condensation of the electromagnetic monopoles we use
the percolation probability $\Pi(A)$. It is the probability that two
infinitely distant points are connected by a monopole cluster (for more
details of the definition see, for example, \cite{BVZ1999}). We found
that this probability is an order parameter, which feels the transition
(Fig.~$3$).  We define the position of the transition using maximum of
the susceptibility $\chi = \langle H^2 \rangle - \langle H\rangle^2$.
It coincides with the point, where the percolation probability
vanishes. At the same time there is no abrupt change of the action on
the transition line. The correlation lengths extracted from the
space-like correlators (\ref{cor}) on the asymmetric lattices do not
increase when approaching the transition line between the phases II and
III.  Let us remind here again that in the conventional lattice
Standard Model the similar transition is found to be a crossover
\cite{EW_T}. However, in \cite{Chernodub} it was found that the
$Z$-strings are condensed at high temperatures in the $3D$ $SU(2)$
Higgs model, while at low temperatures they are not. For this reason
this transition was called in \cite{Chernodub} ``percolation
transition". We did not investigated in detail the transition between
phases II and III in our model. Therefore, we cannot draw a definite
conclusion as to the nature of this transition. However, the
percolation properties of Nambu monopoles show that the vacuum
structure at $T>T_c$ differs from the vacuum structure at $T<T_c$.
Therefore, we do not exclude, that this is actually a phase transition
of a high order.

Investigation of different sizes of the lattices (from $8^3\times 2$ up
to $24^3\times 4$) shows that neither the density nor the percolation
of Nambu monopoles decrease with increasing lattice size for the values
of the couplings considered (up to $\beta = 2.0$) in phase II of the
model. Therefore we suppose that those monopoles survive in the
continuum theory like the Abelian projected monopoles of pure
nonabelian gauge models.

The properties of quantum Cho-Maison monopoles are sufficiently
different from those of the Nambu monopoles as the hypercharge monopole
density within the physical phases II and III decrease rapidly when
moving away from the vertical phase transition line. For this reason we
do not exclude that Cho-Maison monopoles may completely disappear in
the continuum theory.

To conclude, we have considered the $Z_6$ symmetric lattice version of
the Standard Model on the lattice at finite temperature. We must
mention, that the considered lattice model is obtained as a result of
several simplifications.  First of all, we neglect dynamical fermions.
Next, we froze radial fluctuations of the scalar field. Finally, our
choice of gauge couplings corresponds to the $SU(5)$ unified theory. As
a result, strictly speaking, the model may describe color fields
quantitatively only at the energies close to the GUT scale.  However,
we expect that even at the Electroweak scale this model may describe
qualitatively the influence of the emergence of $Z_6$ symmetry in the
lattice action on the phase diagram. We must remind here, that without
$Z_6$ symmetry there is no influence of color fields on the Electroweak
dynamics at all (if one neglects fermion loops).

We have found that at least the lattice model itself differs from the
usual realization of the Standard Model on the lattice. Namely, there
is a confinement-deconfinement phase transition line common to $SU(2)$,
$SU(3)$, and $U(1)$ fields. This line and the line of the transition
which corresponds to the finite temperature electroweak
transition/crossover, meet together in a triple point. However, we do
not consider properly the limit of vanishing lattice spacing. Therefore
at the present moment we do not exclude that the mentioned new
features of the $Z_6$ invariant lattice Standard Model may disappear in
the continuum limit.

Nambu monopoles appeared to be condensed at $T>T_c$ in our lattice
model. We suppose that at $T>T_c$ they survive in the continuum theory.
Finally, if those monopoles are indeed present at $T>T_c$ it would be
natural to suppose, that they may appear at low temperatures in the
form of ordinary particles \footnote{In the original paper by Nambu
\cite{Nambu} the mass of the classical Nambu monopoles was roughly
estimated to be in the TeV region. Moreover, these object were
considered to be confined by the $Z$-string. Therefore, they actually
appear through the bound state composed of a monopole - antimonopole
pair.}. It is worth mentioning, that although we started from the $Z_6$
invariant form of lattice Standard Model, our prediction of the
behavior of quantum Nambu monopoles could be unchanged even without
taking care of $Z_6$ symmetry. This should be, of course, the subject
of another research.

We are grateful to F.V. Gubarev, M.N.Chernodub, E.M. Ilgenfritz, and D.
Boer for useful discussions. A.I.V.  and M.A.Z.  kindly acknowledge the
hospitality of the Department of Physics and Astronomy of the Vrije
Universiteit, where part of this work was done.  This work was partly
supported by the Netherlands Organisation for Scientific Research, by
RFBR grants 06-02-16309, 05-02-16306, and 04-02-16079, RFBR-DFG grant
06-02-04010, by Federal Program of the Russian Ministry of Industry,
Science and Technology No 40.052.1.1.1112, by Grant for leading
scientific schools 843.2006.2.

\clearpage

\begin{figure}
\begin{center}
\epsfig{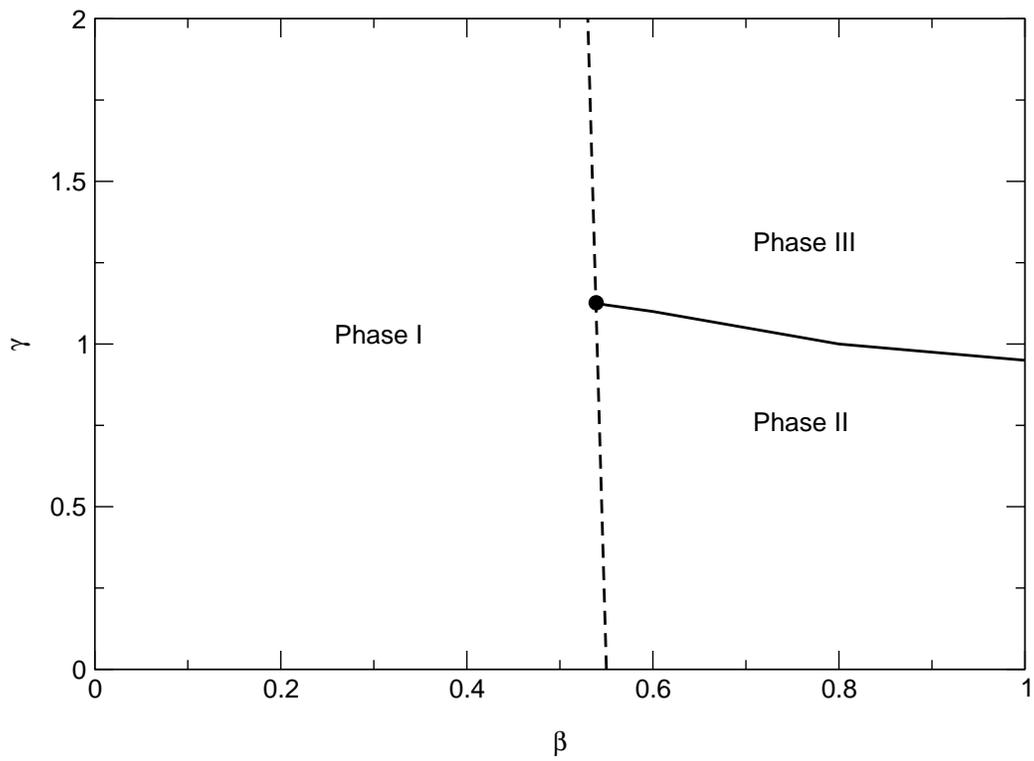}

\caption{\label{fig.1}The phase diagram of the model in the  $(\beta,
\gamma)$-plane for the time extent $N_T = 2$.}

\end{center}
\end{figure}

\begin{figure}
\begin{center}
\epsfig{figure=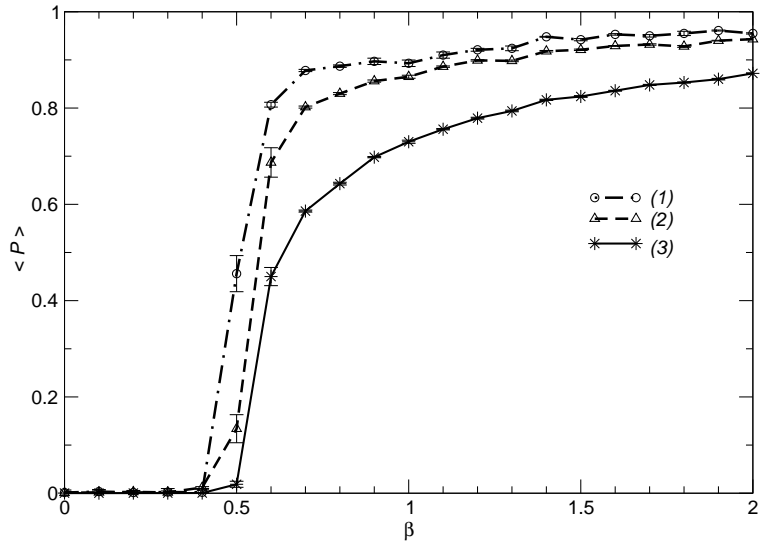,height=100mm,angle=-90}

\vspace{4ex}

\caption{\label{fig.2}The expectation value of Polyakov lines as a
function of $\beta$ for a fixed value $\gamma = 0.5$. (1) corresponds to ${\cal
P}^{\rm R}_{\rm lept}$; (2) corresponds to ${\cal P}^{\rm L}_{\rm lept}$; (3)
corresponds to ${\cal P}^{\rm L}_{\rm quarks}$.}

\end{center}
\end{figure}

\begin{figure}
\begin{center}
\epsfig{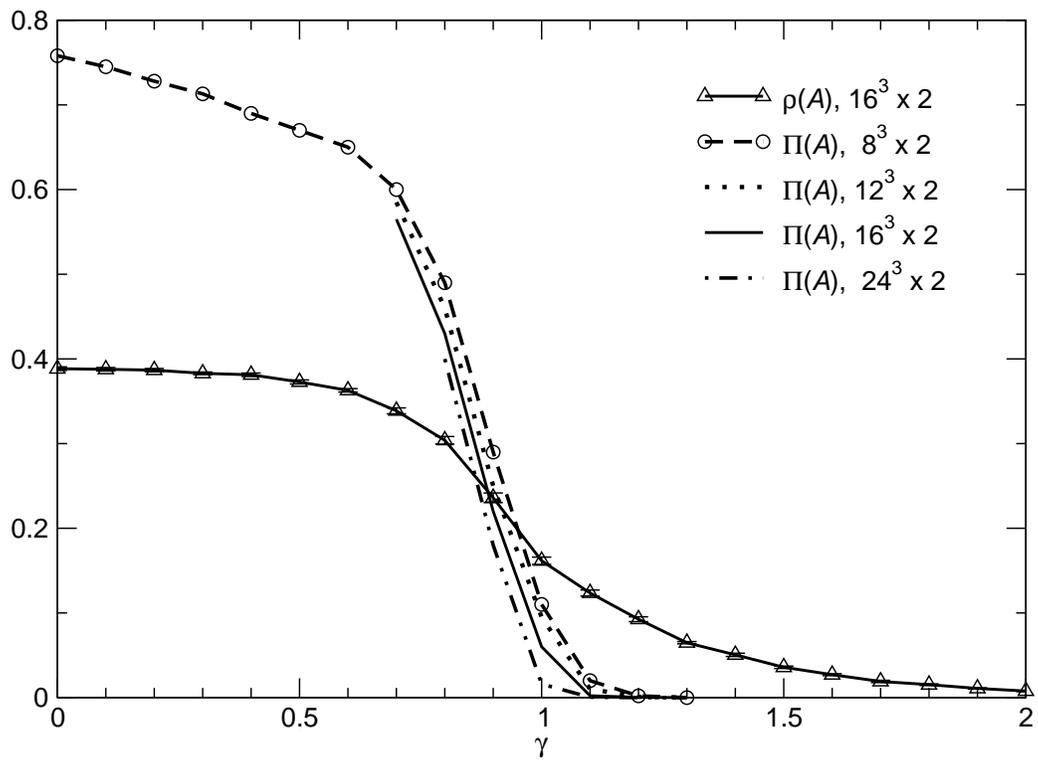}

\vspace{3ex} \caption{\label{fig.3}The density $\rho(A)$ of Nambu monopoles and
the percolation probability $\Pi(A)$ as a function of $\gamma$ for a fixed
value $\beta=0.6$.}

\end{center}
\end{figure}

\end{document}